\documentclass[reprint,superscriptaddress,amsmath,amssymb,aps,pra,twocolumn]{revtex4-2}

\usepackage{subfigure}
\usepackage{tikz}
\usetikzlibrary{calc}
\usepackage[colorlinks,linkcolor=blue,anchorcolor=blue,citecolor=blue]{hyperref}
\usepackage{amsthm, bm, amsmath, braket, multirow}
\usepackage{longtable, array}

\hypersetup{%
    plainpages=true,
    breaklinks=true,
    hypertexnames=false,
    pageanchor=true,
    colorlinks=true,
    linkcolor={blue},
    citecolor={blue},
    urlcolor={red},
    anchorcolor={black}
}
\usepackage{listings, xcolor}
\renewcommand{\arraystretch}{1.2}  

\newcommand{\casql}{Key Laboratory of Quantum Information, Chinese Academy of Sciences, School of Physics, University of Science and Technology of China, Hefei, Anhui, 230026, P. R. China}
\newcommand{\casex}{CAS Center For Excellence in Quantum Information and Quantum Physics, University of Science and Technology of China, Hefei, Anhui, 230026, P. R. China}
\newcommand{\aihf}{Institute of Artificial Intelligence, Hefei Comprehensive National Science Center, Hefei, Anhui, 230088, P. R. China}

\newcommand{\origin}{Origin Quantum Computing, Hefei, Anhui, 230026, P. R. China}
\newcommand{\iat}{Institute of Advanced Technology, University of Science and Technology of China, Hefei, Anhui, 230031, P. R. China}

\begin{document}
\title{SparQSim: Simulating Scalable Quantum Algorithms via Sparse Quantum State Representations}

\author{Tai-Ping Sun}
\affiliation{\casql}
\affiliation{\casex}

\author{Zhao-Yun Chen}
\email{chenzhaoyun@iai.ustc.edu.cn}
\affiliation{\aihf}

\author{Yun-Jie Wang}
\affiliation{\iat}

\author{Cheng Xue}
\affiliation{\aihf}

\author{Huan-Yu Liu}
\affiliation{\casql}
\affiliation{\casex}

\author{Xi-Ning Zhuang}
\affiliation{\casql}
\affiliation{\casex}
\affiliation{\origin}

\author{Xiao-Fan Xu}
\affiliation{\casql}
\affiliation{\casex}

\author{Yu-Chun Wu}
\affiliation{\casql}
\affiliation{\casex}
\affiliation{\aihf}

\author{Guo-Ping Guo}
\affiliation{\casql}
\affiliation{\casex}
\affiliation{\aihf}



\begin{abstract}

    Efficient simulation of large-scale quantum algorithms is pivotal yet challenging due to the exponential growth of the state space inherent in both Sch\"odinger-based and Feynman-based methods. 
    While Feynman-based simulators can be highly efficient when the quantum state is sparse, these simulators often do not fully support the simulation of large-scale, complex quantum algorithms which rely on QRAM and other oracle-based operations. 
    In this work, we present SparQSim, a quantum simulator implemented in C++ and inspired by the Feynman-based method. 
    SparQSim operates at the register level by storing only the nonzero components of the quantum state, enabling flexible and resource-efficient simulation of basic quantum operations and integrated QRAM for advanced applications such as quantum linear system solvers. 
    In particular, numerical experiments on benchmarks from QASMBench and MQTBench demonstrate that SparQSim outperforms conventional Schrödinger-based simulators in both execution time and memory usage for circuits with high sparsity. 
    Moreover, full-process simulations of quantum linear system solvers based on a discrete adiabatic method yield results that are consistent with theoretical predictions. 
    This work establishes SparQSim as a promising platform for the efficient simulation of scalable quantum algorithms.
\end{abstract}

\maketitle


\section{Introduction}\label{sec:intro}

Quantum computing has been extensively studied and has demonstrated potential for quantum advantage in various domains, including linear algebra, machine learning, cryptography, and physical simulations~\cite{Harrow2009,Costa2022,Rebentrost2014,Biamonte2017,Schuld2019,Bennett2014,Google2019,Kim2023,Deng2023}.
However, due to current hardware limitations, realizing large-scale quantum algorithms on quantum computers remains a significant challenge in the NISQ era~\cite{Preskill2018}. 
Simulating quantum algorithms on classical computers is crucial for understanding their behavior and performance, such as validating computational complexity and examining the non-analytical properties of these algorithms~\cite{Huang2020,Markov2018}. 
Moreover, as near-term quantum computers rapidly scale to sizes that challenge direct simulation on classical machines due to memory constraints, efficient simulation techniques for large qubit systems are essential for calibrating and benchmarking both quantum algorithms and hardware~\cite{Google2018}.

Existing simulators primarily rely on the Schr\"odinger and Feynman-based methods, each offering distinct advantages in certain scenarios~\cite{Aaronson2016}. 
The Schr\"odinger-based approach maintains the complete state vector and updates it after each evolution step, whereas the Feynman-based method processes individual state branches independently and updates the state under given quantum operations. 
Additionally, hybrid models that balance these two methods have been proposed~\cite{Aaronson2016,Burgholzer2021,Westrick2024}. 
Although numerous implementations in various programming languages have been developed and parallelism has been achieved using CPUs and GPUs~\cite{Smelyanskiy2016,Haner2017,Markov2018,Pednault2019,Zulehner2019,Huang2020,Markov2020,Villalonga2020,Burgholzer2021,Fatima2021,Suzuki2021,Zhang2021,Jaques2022,Westrick2024}, the simulation of specific quantum algorithms remains a challenging problem.
In particular, the sparsity of state branches can be exploited to enhance the efficiency of Feynman-style simulators. 
Furthermore, incorporating quantum random access memory (QRAM)\cite{QRAM1,QRAM2,QRAM3} and oracle-based implementations could further improve simulation efficiency, though these approaches have been less explored in existing studies.

In this paper, we propose a Feynman-inspired simulator that leverages the sparsity of quantum state branches to simulate quantum algorithms with low memory overhead using C++.
We term this the Sparse Quantum simulator (SparQSim). 
SparQSim is designed to operate at the register level, storing only the nonzero components of the quantum state. 
Its core operations include quantum arithmetic~\cite{QArith1,QArith2,QArith3,QArith4}, basic quantum operations, a comprehensive QRAM implementation that can be seamlessly integrated into quantum algorithms such as quantum linear-system solvers (QLSS)~\cite{Harrow2009,Childs2017,Subasi2019,Costa2022}, with each part being simulated at the register level.
Our benchmarks on circuits from QASMBench~\cite{Li2023} and MQTBench~\cite{Quetschlich2023} indicate that SparQSim performs comparably to mainstream simulators on general benchmarks. 
Numerical experiments further reveal that by exploiting state sparsity, SparQSim outperforms Schr\"odinger-based simulators in both time and memory usage. 
Moreover, SparQSim enables rapid full-process simulations of quantum algorithms such as QLSS using integrated QRAM, with numerical results consistent with theoretical predictions. 
Overall, SparQSim offers a flexible and efficient platform for quantum algorithm simulations, opening new avenues for in-depth research in quantum computing.

\section{Efficient simulation on the register level\label{sec:classical}}

\begin{figure*}[!t]
    \includegraphics[width=\linewidth]{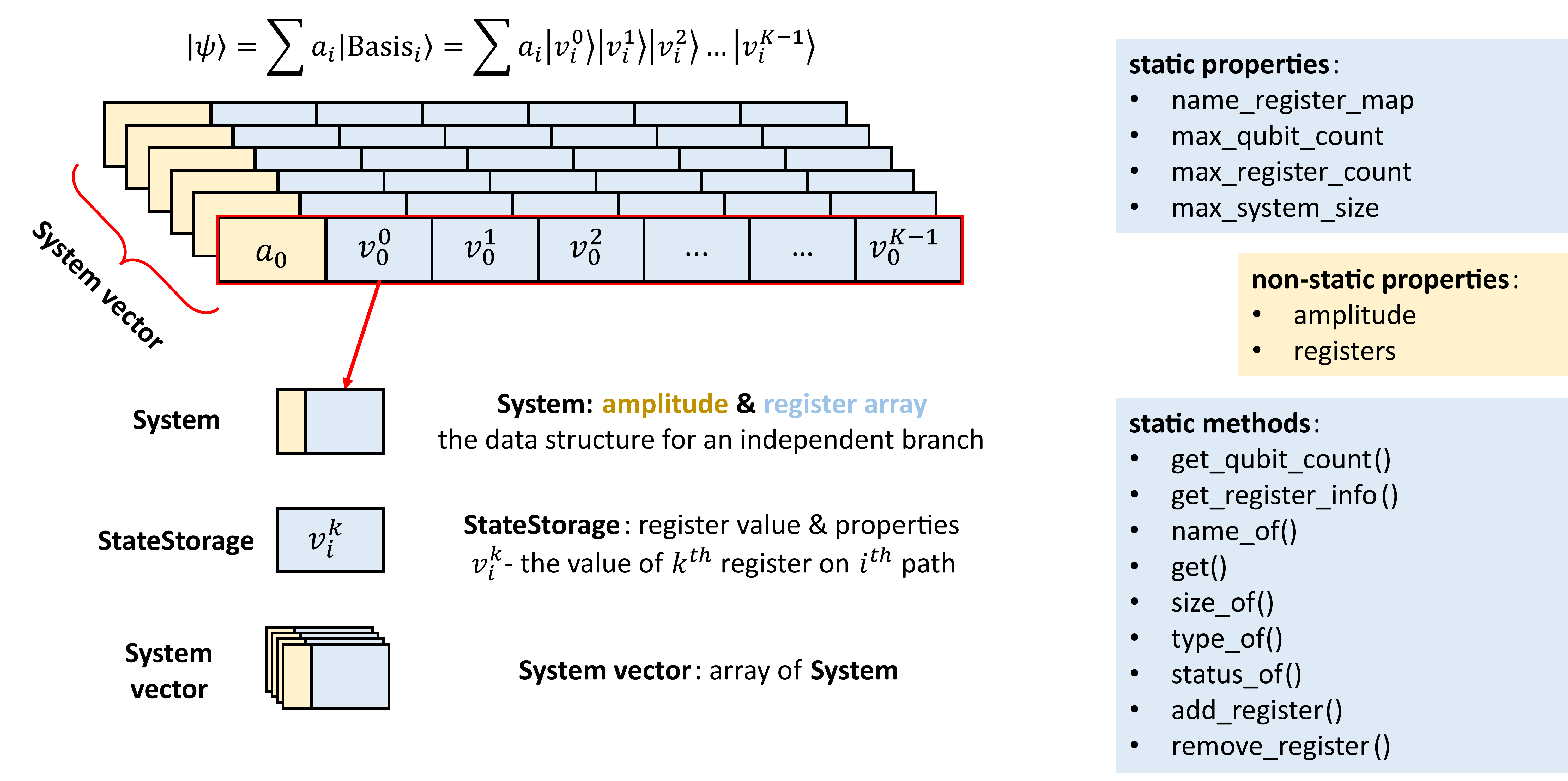}
    \caption{\textbf{Schematic Diagram of Sparse State Representation and Associated Methods.} The left part illustrates the hierarchical data structure for sparse state representation. Each \textbf{System} instance, representing an individual branch (also known as basis), comprises a head amplitude and an ordered array of \textbf{StateStorage} instances. Each element of the array corresponds to a quantum register; for example, $v_i^k$ denotes the value of the $k^{\text{th}}$ register in the $i^{\text{th}}$ branch. The \textbf{System} class includes static properties, non-static properties, and methods that ensure accurate manipulation of each branch. The right panel lists these properties and methods, which manage system-level metadata and facilitate the update of static information for the entire system.}
    \label{fig:system}
\end{figure*}

\subsection{Qubits and quantum operations}
\textbf{Qubits.} A quantum state represents the state of a quantum system, often described using wave functions or state vectors. 
In the realm of quantum computing, these vectors can be written in the form of a superposition of basic states of the corresponding Hilbert space: 
\begin{equation}
    \ket{\psi} = \sum_i c_i \ket{i},
    \label{eq:state}
\end{equation}
where $c_i\in \mathbb{C}$ and $\ket{i}$ denotes computational basis. 

A qubit (quantum bit) is the basic unit of quantum information and quantum computing, similar to a bit in classical computing. 
However, unlike a classical bit, which is either in state 0 or 1, a qubit can be in a superposition of both states. 
A single qubit state can be represented as:
\begin{equation}
    \ket{\psi} = \alpha \ket{0} + \beta \ket{1},
\end{equation}
where $\alpha, \beta \in \mathbb{C}$ are the amplitudes of the corresponding basis and satisfy normalization condition $ |\alpha|^2 + |\beta|^2 = 1$. 
Since $\ket{0}$ and $\ket{1}$ are orthogonal and span the space $\mathbb{C}^2$, the state of a qubit can be written as a linear combination of these two bases:
\begin{equation}
    \ket{\psi} = \alpha \begin{bmatrix} 1 \\ 0 \end{bmatrix}  + \beta \begin{bmatrix} 0 \\ 1 \end{bmatrix} . 
    \label{eq:qubit_state}
\end{equation}
With this qubit representation, each basis in $n$-qubit system of Eq.~(\ref{eq:state}) is actually a Kronecker product of $n$ qubits: 
\begin{equation}
    \ket{i}:=\ket{i_{n-1}}\otimes \cdots \otimes \ket{i_0}
\end{equation}
where $i_k$ denotes the $k$-th bit of the integer $i$. 

\textbf{Quantum operations.} Quantum operations, or quantum gates, manipulate qubits and alter quantum states. 
They are analogous to logical operations in classical computing but work differently due to quantum superposition and entanglement. 
A quantum operation $U$ that acts on $n$ qubits satisfies the unitary condition, $U^\dagger U=UU^\dagger=I$, where $U^\dagger$ represents the Hermitian conjugate of $U$. 
For consistency with quantum states' vector formulation, the operation $U$ is typically represented as a $2^n\times 2^n$-dimensional complex matrix that act on state vectors. 
For example, the single-qubit Pauli-X gate, denoted as $X$ and functioning as the classical NOT gate, is represented as $X=\begin{bmatrix} 0 & 1 \\ 1 & 0 \end{bmatrix}$. 
Other common single-qubit operations include $Y$, $Z$, and $H$, which are defined as $Y=\begin{bmatrix} 0 &-i\\ i&0\end{bmatrix}$, $Z=\begin{bmatrix} 1 & 0 \\ 0 & -1 \end{bmatrix}$, and $H=\frac{1}{\sqrt{2}}\begin{bmatrix} 1 & 1 \\ 1 & -1 \end{bmatrix}$. 
The two-qubit operations, typically the controlled single-qubit operations, can be represented as $CU=\ket{0}\bra{0}\otimes I+\ket{1}\bra{1}\otimes U$ if the control qubit is $1$. 
Quantum operations serve as the fundamental building blocks for the quantum circuit model, and simulating the behavior of these operations on quantum states forms the foundation for the simulation of quantum algorithms.

\subsection{Motivation} \label{sec:motiv}

Simulating quantum algorithms is a computationally challenging task due to the exponential growth of the Hilbert space as the number of qubits increases. 
This growth leads to exponential scaling in either time or space complexity~\cite{Aaronson2016}.
Current quantum simulators are primarily based on the Schr\"odinger method~\cite{Haner2017,Google2018,Google2019,Zulehner2019} or the Feynman-based method~\cite{Aaronson2016,Markov2018,Westrick2024}. 
Schr\"odinger-based simulators store and update a full state vector at each quantum gate, resulting in time complexity $\mathcal{O}(m2^n)$ and space complexity of $\mathcal{O}(2^n)$, where $n$ is the number of qubits and $m$ is the number of gates.
In contrast, the Feynman-based method reduces space complexity to $\mathcal{O}(m+n)$ by applying $m$ gates to individual branches and summing over them with non-zero amplitudes, which are known as Feynman paths, but it incurs a time complexity of $\mathcal{O}(4^m)$. 
When quantum states exhibit sparsity—where the number of nonzero paths is significantly smaller than the total number of paths—the Feynman-based approach becomes more efficient than the Schrödinger-based method.

Although some quantum simulators leverage state sparsity~\cite{Jaques2022,Westrick2024}, they typically lack support for simulations that integrate oracles with quantum circuits, such as data input oracles, QRAM, and quantum adder oracles—key components in quantum arithmetic operations. 
Integrating these oracles into simulations is crucial for facilitating resource estimation~\cite{Dalzell2023}, a growing area of research that aims to demonstrate the potential of quantum advantages, especially in the context of early fault-tolerant quantum computing. 
On the other hand, the development of new quantum algorithms and their applications increasingly requires fast numerical results. 
In these cases, oracles like quantum arithmetic operations can be simulated as a single unit, where the input-output mapping is more important than the actual quantum circuit, which involves complex transpiling and optimization. 
Given these limitations, there is an increasing demand for more flexible and full-process simulations of quantum algorithms, such as the QLSS, which can benefit from an integrated approach to oracles.

In response to these challenges and limitations, we introduce SparQSim, a modified Feynman-based simulator that leverages state sparsity in quantum systems and supports the integration of oracles, including QRAM. This approach overcomes the limitations of existing simulators on certain scenarios and broadens their applicability to a wider range of quantum algorithms.

\subsection{Sparse state representation and data structure}

In Feynman-based simulation, the approach involves tracking the evolution of a quantum state by summing over all possible weighted paths. 
As quantum operations are applied to the initial state, the branches, which represents computational basis states at each evolution step, proliferate in a tree-like structure, with the branch count peaking at the root and evolving as operations proceed. 
A straightforward implementation of this method leverages the binary representation of the basis states by decomposing them into individual qubits. 
However, this approach hinders efficient addressing of a group of qubits, which are the target of a quantum operation, such as an oracle. 
To address this, we propose a register-level sparse state representation, in which each register is represented as a higher-level quantum system in bra-ket notation. 
This representation stores only the quantum branches with non-zero amplitudes, along with the corresponding register metadata for each branch. 
A quantum state can thus be written as:
\begin{equation}
    |\psi\rangle = \sum_i a_i|v^0_i\rangle|v^1_i\rangle|v^2_i\rangle...|v^{K-1}_i\rangle,
\end{equation}
where $v^j_i$ represents a computational basis state of register $j$ in the path $i$, and $K$ is the number of registers. 

Figure~\ref{fig:system} shows a schematic of the data structure.
To efficiently manipulate each branch, we define the data structure of the \textbf{System} and its properties/methods. 
As shown in Fig.~\ref{fig:system}, the \textbf{System} class contains the amplitude and register metadata array, where each register is an instance of \textbf{StateStorage} class. 
Different \textbf{System} instances form the \textbf{System} vector, which represents the quantum state. From the length we can get a basic idea of the state's sparsity. 

The manipulation of the quantum state, represented as a \textbf{System} \textbf{vector}, is performed by retrieving and updating the metadata of each branch's registers accordingly. 
Each register's information is stored as a plain binary string, with its length corresponding to the number of qubits in the register. 
In practice, for memory efficiency, all elements are represented by fixed 64-bit binary strings, with the actual length truncated to match the number of qubits.
To create a new register, an element in the register array is activated; conversely, to remove a register, the corresponding element is simply deactivated. Both the manipulations are achieved by static methods of the \textbf{System} class. 

The key properties and methods of the \textbf{System} and \textbf{StateStorage} classes, which manage the metadata and update the static information of the entire system, are listed in Fig.~\ref{fig:system}. 
For example, static properties of \textbf{System} such as \verb|name_register_map|, \verb|max_qubit_count|, \verb|max_register_count| and \verb|max_system_size| provide not only information about each register but also details about the current system size and qubit count. 
Non-static properties, such as \verb|amplitude| and \verb|registers|, store the amplitude of each branch and the actual data of the registers. 
By utilizing these properties and methods, we can easily manipulate each branch of the system. 
These mechanisms ensure information consistency and enable global information sharing among all branches. 
Tables~\ref{tb:pm} and \ref{tb:pm2} in the supplementary material list the properties and methods of the \textbf{System} and \textbf{StateStorage} classes, along with their brief descriptions.

\subsection{Classical simulation of the quantum operations}
\subsubsection{Non-interference operations}

Non-interference operations are those that act on each computational branch independently, meaning that each branch is processed without interacting the others.
Specifically, these operations involve either phase modification on certain branches or bit flips on specific qubits.

\begin{figure}[!h]
    \includegraphics[width=\linewidth]{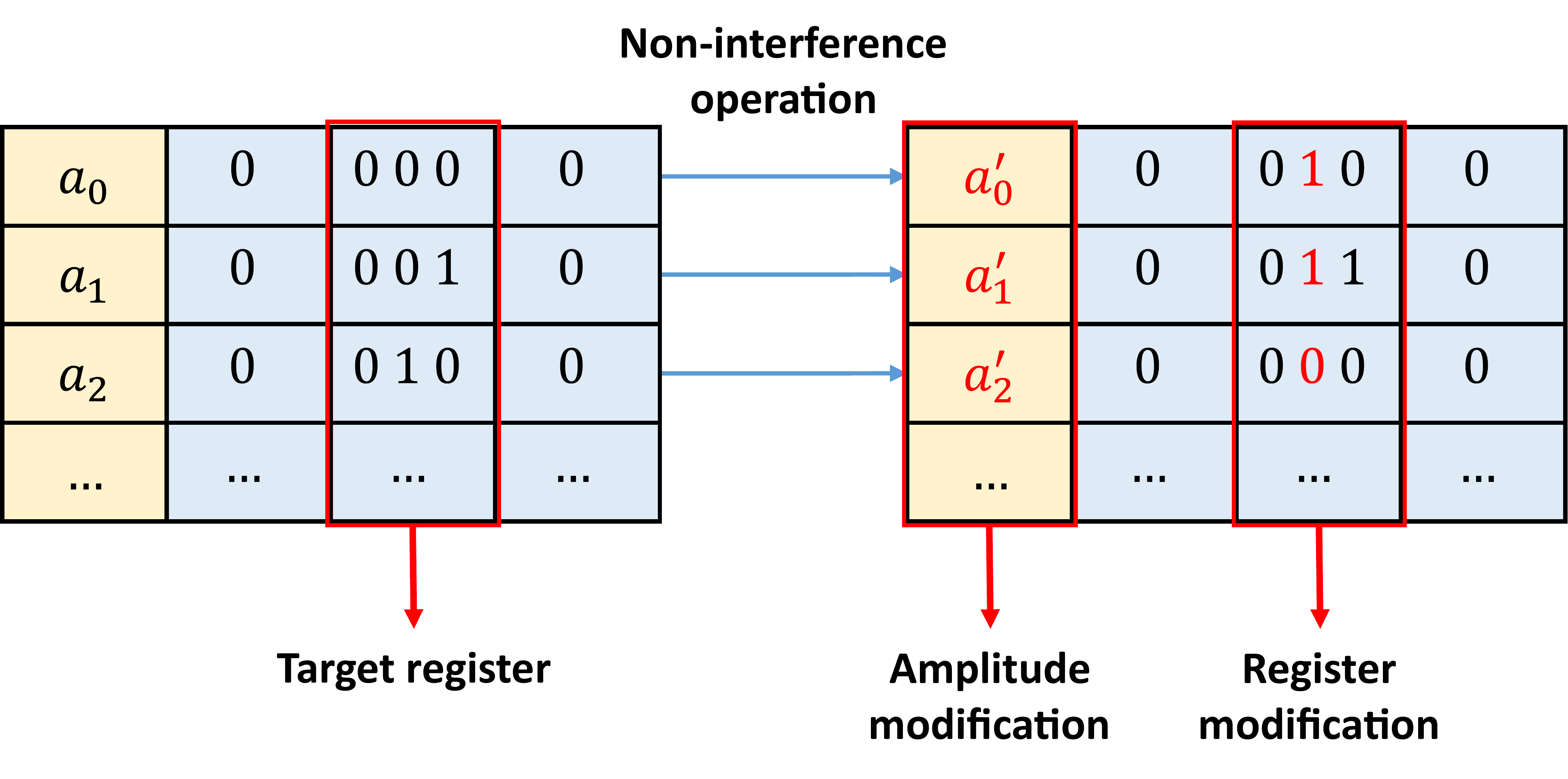}
    \caption{\textbf{Non-interference Operations.} A non-interference operation transforms the system vector by independently modifying the amplitude and register values. The modified segments are highlighted in red.}
    \label{fig:non-interference}
\end{figure}

The register-level simulation of non-interference operations follows these basic steps, as shown in Fig.~\ref{fig:non-interference}. First, iterate over all branches and, for each branch, extract the head value of the \textbf{System}, the amplitude, and the contents of the input registers. Next, process the extracted values with rules of corresponding operations to obtain new temporary values. Finally, update these values of the amplitudes and the contents in the branch.
It is important to note that the modification of amplitudes primarily involves phase changes, which are represented by multiplying with a complex phase factor. 
Therefore, non-interference operations do not require zero-branch clearing or changes in the number of branches.
This approach also facilitates the controlled version of these operations. 
To implement this, the control qubit value can be extracted, and an if-else statement can be added during the branch iteration step to apply the operation conditionally.

A critical aspect of these operations is ensuring that the changes to the register contents are reversible. 
Single-qubit operations such as Pauli gates $\{X, Y, Z\}$ and phase gates $\{S, T\}$, as well as their controlled counterparts, are typical examples of non-interference operations. 
Their inverse operations can be easily implemented by flipping the bits and modifying the amplitudes. 
In addition to these reversible single-qubit and controlled operations, another important class of operations in quantum computing is quantum arithmetic, which encompasses operations such as addition, multiplication, and division~\cite{QArith1}. 
These operations are implemented by the same method and serve as quantum oracles without explicit circuit transpiling and optimization.

\subsubsection{Interference operations}

\begin{figure*}[!t]
    \includegraphics[width=\linewidth]{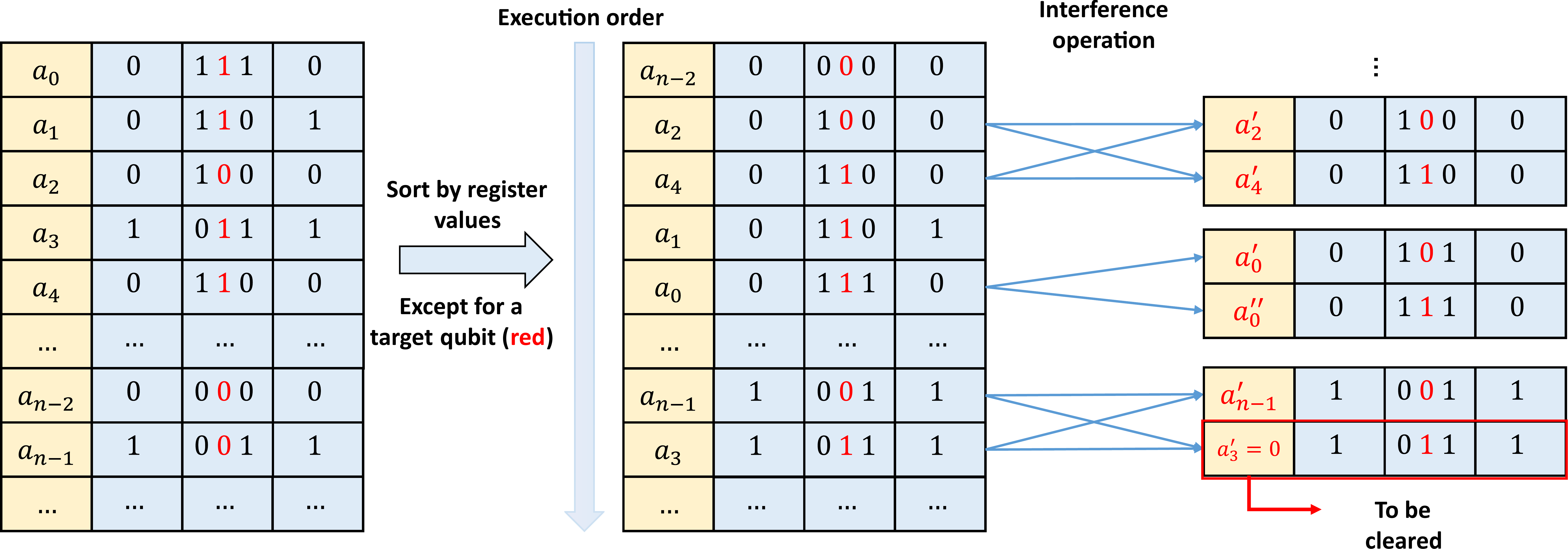}
    \caption{\textbf{Interference Operations.} The state vector is first sorted according to the idle register values, with the target qubit (the operated bit) highlighted in red. This sorted order ensures that coherent pairs corresponding to the target qubit appear consecutively. Next, the interference operation is applied to both individual and coherent components. Finally, branches with zero amplitudes are removed for further computation.}
    \label{fig:interference}
\end{figure*}

Interference operations, related to quantum interference, combine coherent branches by either amplifying or diminishing their amplitudes. 
Such operations often result in the creation of new branches or the elimination of existing ones. 
For example, a Hadamard gate transforms the basis state $\ket{0}$ into the superposition $\ket{+} = \frac{1}{\sqrt{2}}\left( \ket{0} + \ket{1} \right)$ and can also revert $\ket{+}$ back to $\ket{0}$.

Suppose the interference operation is applied to a set of qubits, referred to as the target qubits, on corresponding registers; the remaining qubits, which are not directly involved in the operation, are termed idle qubits. 
Branches remain coherent if and only if the values of all idle qubits are identical across those branches.

The first step of the operation is to sort all branches and partition them into groups where each group's elements are coherent, based on the ascending values of the idle qubits. 
The second step is to apply the operation to each group. 
If a group contains a number of branches equal to the dimension of the operation (for instance, a Hadamard gate acting on $k$ qubits will produce $2^k$ branches), the operation can be computed in place. 
Otherwise, additional branches may be appended to ensure the operation is performed across the full dimensionality.

A simplified case occurs when the number of target qubits is 1, and the size of the group elements is at most 2. As illustrated in Fig.~\ref{fig:interference}, the target qubit is highlighted in red. The elements of the execution list are either operated on individually or coherently, which may lead to new branches or the amplification or diminishment of existing branches.

After computation, some branches may acquire zero amplitude due to destructive interference. These branches are then removed from the system vector to prevent unnecessary computational overhead in further steps.


\subsubsection{Multi-threading with OpenMP}
SparQSim is designed to support multi-threading with OpenMP. 
In the case of non-interference operations, the branches of the state remain independent throughout the simulation. 
This independence allows parallelism to be achieved by dividing the branches into several chunks and assigning each chunk to a different thread. 
Each thread can compute the branches within its assigned chunk independently, and then combine the results to update the state.
However, interference operations cannot be computed in parallel in the same way as non-interference operations because the different chunks cannot be processed with the same shared system vector. 
For example, adding a new branch cannot be parallelized due to the risk of data races. 
In interference operations, only the sorting step can be parallelized, which can be achieved using merge sort. 
The OpenMP library provides a simple interface for multi-threading, and the SparQSim can be easily parallelized by adding a few lines of code. 



\section{Results}
In this section, we first provide a detailed analysis of SparQSim's performance on quantum operations and general benchmarks using circuits from QASMBench~\cite{Li2023} and MQTBench~\cite{Quetschlich2023}, with input in QASM format~\cite{Cross2017}. 
We present detailed measurements of time and memory consumption for different operations, comparing SparQSim's performance to other simulators. 
The benchmarked circuits are selected from various quantum algorithms with different qubit counts to assess SparQSim's efficacy. 
The running time and peak memory usage are measured over 10 repeated trials on a Linux system.
Next, we implement the widely used quantum linear system solver algorithm, based on the discrete adiabatic theorem, as a full-process quantum algorithm application~\cite{Costa2022}. 
We provide a detailed numerical analysis of the impact of matrix size and data encoding style on performance.

All experiments and benchmarks are conducted on a 128-threads server equipped with two 3.50 GHz Intel Xeon Platinum 8369B CPUs (each with 64 threads) and 512 GB of total memory. 
A list of libraries and versions used for the CPU benchmarks is provided in Tab.~\ref{tab:libs}.

\begin{table}
    \begin{tabular}{ll}
        \hline
        Library & Version \\
        \hline
        G++ & 11.4.0 \\
        Cmake & 3.22.1 \\
        Ninja & 1.10.1 \\
        MPL & 0.4 \\
        Python & 3.10.8 \\
        Qiskit & 1.2.4 \\
        Qiskit-Aer & 0.15.1 \\
        Qsimcirq & 0.21.0 \\
        Scipy & 1.14.1 \\
        Numpy & 1.26.4 \\
        \hline
    \end{tabular}
    \caption{\textbf{Required Libraries, Compilers, and Packages along with Corresponding Versions Used for Benchmarking.}}
    \label{tab:libs}
\end{table}

\subsection{Performance of basic operations and general benchmarks}\label{sec:result1}

As we observe, state sparsity contributes to the efficiency of quantum algorithms, and this efficiency decays as the number of branches increases. 
Therefore, the bound of efficiency should be determined through numerical simulations of different quantum algorithms. 
Additionally, as a specialized type of simulator, it is necessary to compare SparQSim with existing simulators to highlight its advantages on certain quantum algorithms. 
We emphasize that the goal of this study is not to replace existing simulators with one that consumes fewer resources, such as running time and memory usage, across various quantum algorithms with different levels of state sparsity. 
Instead, our aim is to provide a user-friendly and efficient tool for quantum algorithm development and testing.

\subsubsection{Performance of quantum operations on different sparsity levels} \label{sec:result1-1}

\begin{figure*}[!t]
    \centering
    \begin{tikzpicture}
        \node[anchor=center] (center) at (0,0){
            \includegraphics[width=\linewidth]{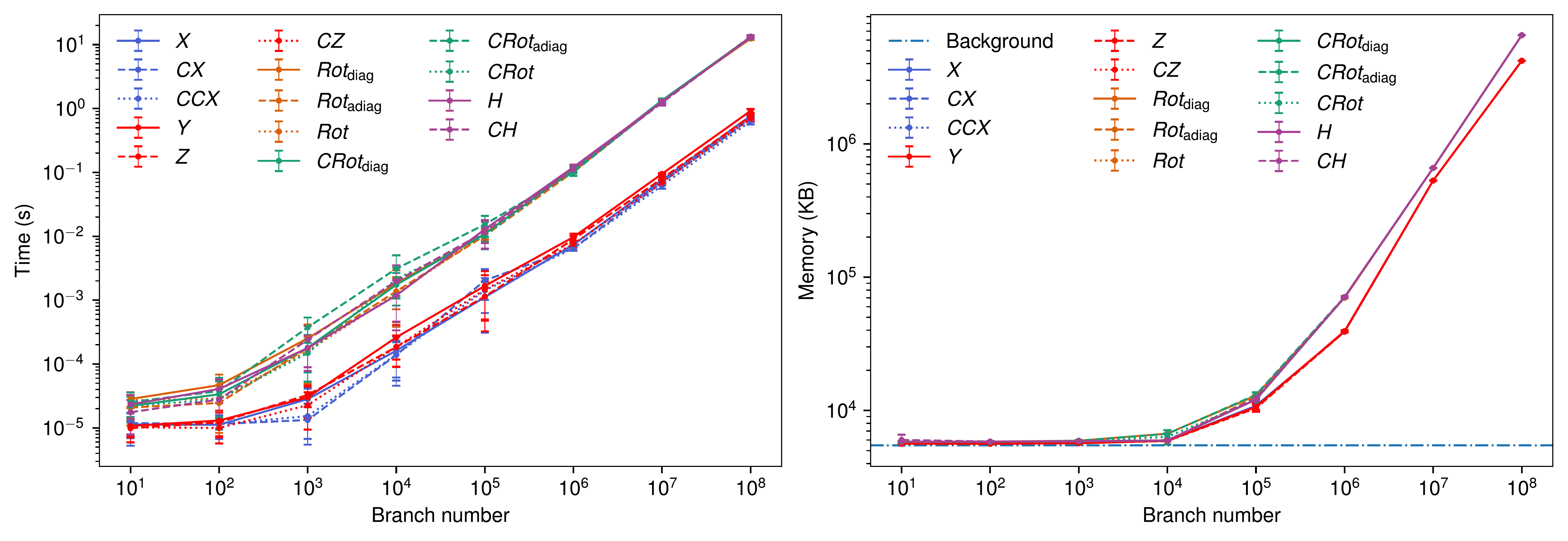}
        };
        \node[anchor=center] (a) at ($(center.north west)+(0.5cm, -0.5cm)$) {\textbf{(a)}};
        \node[anchor=center] (b) at ($(center.north)+(0.5cm, -0.5cm)$) {\textbf{(b)}};
    \end{tikzpicture}
    
    \caption{\textbf{Time Cost and Memory Usage vs. Branch Numbers.} (a) Execution time (s) of non-interference operations is plotted against branch numbers. (b) Peak memory usage (KB) of the simulation is plotted against branch numbers. A reference line indicating the background memory usage for a single branch is included for comparison. Two distinct families of curves are observed in both sub-figures. The lower family, corresponding to non-interference operations (e.g., $X$, $Y$, $Z$ and their controlled versions), exhibits lower time and memory costs, whereas the upper family, corresponding to interference operations (e.g., $Rot$, $H$ and their controlled versions), demonstrates higher resource requirements. For the rotation gate, $Rot_{\text{diag}}$ and $Rot_{\text{adiag}}$ denote the diagonal and anti-diagonal implementations, respectively. Both axes are displayed on a logarithmic scale.}
    \label{fig:sparsity}
\end{figure*}

As a quantum circuit simulator inspired by the Feynman-based simulation, SparQSim benefits from varying levels of state sparsity. 
The higher the sparsity, the more efficient the algorithm becomes. 
In this section, we compare the performance of SparQSim across different sparsity levels on initial branches.

To demonstrate the availability and stability of SparQSim, we test the time cost and memory consumption for different quantum operations, which can be divided into two groups: non-interference operations and interference operations.  
The non-interference group includes single-qubit gates, $\{X, Y, Z\}$, as well as their controlled counterparts $\{CX, CCX, CZ\}$. 
The interference group includes Hadamard gate $H$ and its controlled version $CH$, general rotation gates $Rot, CRot$, and variants of different matrix forms, such as diagonal and anti-diagonal, denoted as $\{Rot_{\text{diag}}, Rot_{\text{adiag}}, CRot_{\text{diag}}, CRot_{\text{adiag}}\}$. 
The initial size of the system vector (i.e., the branch number) is set to range from $10$ to $10^8$, with equal intervals on a logarithmic scale. For each configuration, we run $10$ trials using a single thread, and then compute the average running time and peak memory usage with standard deviation.

In Fig.~\ref{fig:sparsity} (a), the lower-left portion of the curves shows a modest overhead with a slow increase. 
This occurs because when the number of branches is low, the cost of the state representation is relatively minor compared to the system initialization overhead. 
However, when the number of branches surpasses $10^3$, the time cost for all operations demonstrates a linear relationship with the branch number. 
This is consistent with the time complexity $\mathcal{O}(4^m)$ discussed in Sec.~\ref{sec:motiv}. 
While we test a single quantum operation here, the initial number of branches we set implies that at least $m$ interference operations are required to generate a given number of branches, assuming the system starts from a single branch $\ket{0}$. 
Fig.~\ref{fig:sparsity} (a) also shows a clear time distinction between the two groups of operations, with non-interference operations being more time-efficient for the same number of branches due to their characteristic of not creating new branches. 

A similar trend can be observed in Fig.~\ref{fig:sparsity} (b). When the branch number is below $10^4$, the memory usage remains nearly constant, represented by a horizontal line. 
This baseline represents the background memory usage when the branch number is just one, as the least branch number to simulate.
Once the branch number exceeds $10^5$, the memory usage increases approximately linearly with the branch number. 
It is important to note that the memory usage measured here refers to the memory required for the system vector, not the operation itself. 
Therefore, the space complexity is not the $\mathcal{O}(m+n)$ mentioned earlier. 
In other words, the memory usage of SparQSim is primarily driven by the state representation rather than the operation simulation, as the state needs to be stored for further computation.

\subsubsection{Performance of speed-up by multi-threading} \label{sec:result1-2}

\begin{figure}[!t]
    \includegraphics[width=\linewidth]{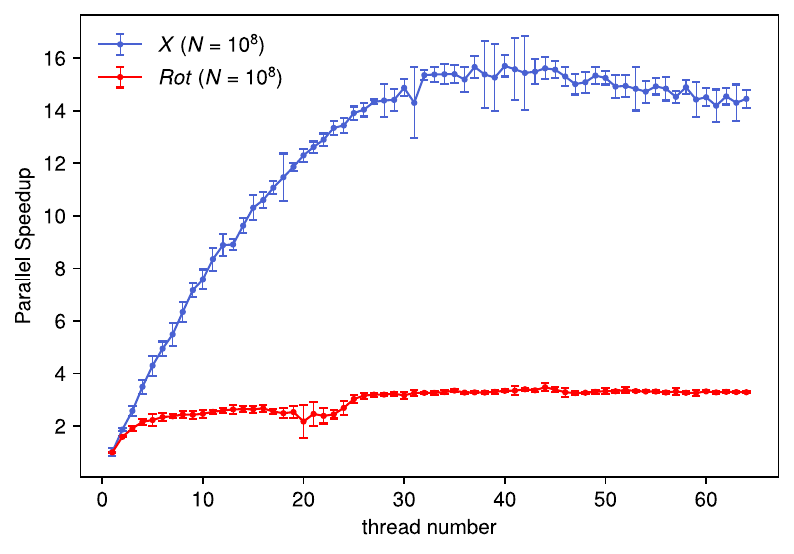}
    \caption{\textbf{Parallel Efficiency of Different Quantum Operations.} A typical non-interference operation $X$ (blue) and a typical interference operation $Rot$ (red) are simulated with thread counts ranging from 1 to 64 to showcase the parallel efficiency defined in Eq.~\ref{eq:speedup}. The branch number is fixed at $10^8$ to minimize state initialization overhead.}

    \label{fig:speedup}
\end{figure}

In this section, we evaluate the performance improvement achieved through parallelization using OpenMP. 
Since interference and non-interference operations adopt different strategies and exhibit distinct time scaling patterns (as shown in Fig.~\ref{fig:sparsity}), our tests of parallelization are conducted separately for these two categories. 
We selected two representative operations—namely, the $X$ gate to represent non-interference operations and a general rotation gate ($Rot$) to represent interference operations—because these single-qubit operations capture the essential characteristics of their respective groups. 
Controlled versions of these gates are not tested, as their performance closely mirrors that of the single-qubit versions while adding additional overhead without substantially altering parallelization behavior.

To facilitate comparison, we define parallel speed-up as the ratio of the execution time on a single thread to that on $p$ threads:
\begin{equation}\label{eq:speedup}
    \text{Speedup}(p)=\frac{T(1)}{T(p)},
\end{equation}
where $T(1)$ and $T(p)$ denote the execution times for single-threaded and $p$-threaded runs, respectively. 
Since the execution time for a low number of branches is very small, variability between trials can lead to inaccurate speed-up measurements. 
To mitigate this effect, we fix the branch number at $10^8$ and average the results over 10 trials. 
The results are shown in Fig.~\ref{fig:speedup}.

From Fig.~\ref{fig:speedup}, the parallel speedup increases gradually with the number of available threads, varying from 1 to 64 as set by the \verb|OMP_NUM_THREADS| variable, for the quantum operations tested. 
Non-interference operations, represented by the $X$ gate, achieve better parallel efficiency compared to interference operations, represented by the $Rot$ gate. 
This is consistent with Amdahl's law, which suggests that the parallel efficiency of the latter is constrained by the time required for the non-parallelizable portion of the task.
As the number of threads reaches approximately $32$, the parallel speedup saturates, and parallelization overhead begins to increase. 
As a result, the parallel efficiency of SparQSim starts to decrease.

Overall, the results demonstrate that SparQSim is able to achieve significant speedup for non-interference operations. 
Meanwhile, the parallel efficiency of interference operations shows room for further optimization in the future.

\subsubsection{Performance of benchmarks to other different simulators} \label{sec:result1-3}

\newcommand{\colwidth}{1.5cm}
\begin{table*}
    \begin{tabular}{|l|l|l|p{\colwidth}|p{\colwidth}|p{\colwidth}|p{\colwidth}|p{\colwidth}|p{\colwidth}|p{\colwidth}|p{\colwidth}|}
        \hline
        \multicolumn{2}{|c|}{Benchmarks} & \multicolumn{1}{c|}{Sparsity} & \multicolumn{4}{c|}{Time (s)} & \multicolumn{4}{c|}{Memory (MB)} \\
        \hline
        types & \# qubits & (\# branches) &SparQSim & Qiskit & Qsim & GraFeyn & SparQSim & Qiskit & Qsim & GraFeyn \\
        \hline
        \multirow{3}{*}{ghz} & 23 &2 &0.0005 &1.393 &0.2339 &0.0001        &5.642 &335.391 &412.099 &7.631        \\
        \cline{2-11}
         & 40 &2 &0.0006 &---&---&0.0001 &5.631 &---&---&7.779         \\
        \cline{2-11}
        & 255 &2 &0.0006 &---&---&0.0010 &6.038 &---&---&11.737 \\
         \hline
        \multirow{1}{*}{qram} & 20 &1 &$<10^{-4}$ &0.1577 &0.0386 &0.0002        &4.814 &221.705 &299.466 &7.723        \\
         \hline
        \multirow{1}{*}{qft} & 29 &536870912 &597.557&589.5122&156.5386&210.7924    &59982.569&8400.235&8479.073&10220.904     \\
        \hline
        \multirow{1}{*}{swap\_test} & 25 &33554432 &14.8518 &6.3044 &0.7921 &7.6762  &4659.411 &715.985 &797.573 &692.869 \\
        \hline
    \end{tabular}
    \caption{\textbf{Experimental Results of SparQSim Compared to Qiskit, QSim, and GraFeyn in Single-Thread Mode.} Running time (in seconds) and memory usage (in MB) are averaged over 10 trials. Cells marked with ``---" indicate either a timeout failure or an out-of-memory error.}
    \label{tab:benchs1}
\end{table*}

\begin{table*}
    
    \begin{tabular}{|l|l|l|p{\colwidth}|p{\colwidth}|p{\colwidth}|p{\colwidth}|p{\colwidth}|p{\colwidth}|p{\colwidth}|p{\colwidth}|}
        \hline
        \multicolumn{2}{|c|}{Benchmarks} & \multicolumn{1}{c|}{Sparsity} & \multicolumn{4}{c|}{Time (s)} & \multicolumn{4}{c|}{Memory (MB)} \\
        \hline
        types & \# qubits & (\# branches) &SparQSim & Qiskit & Qsim & GraFeyn & SparQSim & Qiskit & Qsim & GraFeyn \\
        \hline
        \multirow{3}{*}{ghz} & 23 &2 &0.0006&0.2736&0.3628&0.0003&5.601&333.23&411.237&174.529        \\
        \cline{2-11}
         & 40 &2 &0.0005 &---&---&0.0002 &5.622 &---&---&187.149         \\
        \cline{2-11}
        & 255 &2 &0.0008 &---&---&0.0015 &6.041 &---&---&185.389      \\
         \hline
        \multirow{1}{*}{qram} & 20 &1 &0.4743&0.429&0.3007&0.0002 &7.008&220.17&295.215&172.07     \\
         \hline
        \multirow{1}{*}{qft} & 29 &536870912 &106.1023&22.8111&30.517&6.1448 &59987.13&8408.472&8483.344&9264.565        \\
        \hline
        \multirow{1}{*}{swap\_test} & 25 &33554432 &5.6675 &0.3681 &0.3818 &0.3282   &3946.825 &722.668 &796.813 &1003.785        \\
        \hline
    \end{tabular}
    \caption{\textbf{Experimental Results of SparQSim Compared to Qiskit, QSim, and GraFeyn with 64 Threads.} The settings are the same as in Table~\ref{tab:benchs1}, but simulations are executed using 64 threads.}
    \label{tab:benchs3}
\end{table*}

In this part, we evaluate the performance of SparQSim by comparing it with several widely used quantum simulators, including Qiskit, Qsim, and GraFeyn. 
To assess SparQSim's efficiency, particularly in cases where quantum circuits exhibit significant sparsity, we conduct benchmarks across various circuit groups. 
These simulators represent two simulation paradigms: Schr\"odinger-based simulators (Qiskit and Qsim) and Feynman-based simulators (GraFeyn).

A typical group of circuits consists of those with high sparsity, such as GHZ states with varying qubit numbers, and QRAM circuits (with single-branch input). 
These circuits involve relatively few interference operations, and the branch number remains relatively small, making them ideal for testing SparQSim's performance. 
Another group of circuits consists of those with low sparsity, such as Quantum Fourier Transform (QFT) and Swap Test circuits, where the peak branch number approaches 100\% of the total number of state branches. 
This increases the memory requirement and time consumption of circuit simulation to $\mathcal{O}(2^n)$, which diminishes the advantages of SparQSim. 
Testing these circuits is important for evaluating SparQSim's performance when handling large-scale quantum circuits with large branch numbers.
We execute the benchmark circuits and record both the running time and peak memory usage. 
Based on multi-threading availability, we test the different simulators using both single-thread and 64-thread parallelism. 
The results are shown in Tables~\ref{tab:benchs1} and~\ref{tab:benchs3}.

In both single-threaded and multi-threaded scenarios, SparQSim outperforms Schrödinger-based simulators (Qiskit and Qsim) in terms of time efficiency, especially for circuits with sparse state representations. 
It demonstrates similar performance to the hybrid model simulator, GraFeyn, on sparse circuits. 
However, when the branch number increases and the state becomes denser, SparQSim's time cost exceeds that of the other three simulators, as SparQSim is not optimized for dense circuits.

In terms of memory usage, SparQSim shows a significant advantage in the sparse case, using considerably less memory compared to the other simulators. 
However, as the branch number increases and the state becomes dense, SparQSim's memory usage also increases and can surpass that of the other simulators.
Notably, we observed that multi-threading does not significantly impact SparQSim's memory usage, unlike GraFeyn, where memory usage increases considerably with parallelization.


\subsection{A full-process implementation of the quantum linear-system solver} 
\label{sec:result2}

Quantum linear-system problems are ubiquitous in many applications, such as solving differential equations~\cite{Berry2014,Berry2017}, data fitting~\cite{Wiebe2012}, and machine learning~\cite{Rebentrost2014}, among others. 
Specifically, the goal of a quantum linear-system solver is to generate a solution encoded into a quantum state $\ket{\bm{x}}$ that satisfies $\|\ket{\bm{x}}-\tilde{\bm{x}}\|<\epsilon$, where $\epsilon$ is a given precision and $\tilde{\bm{x}}=\bm{x}/\|\bm{x}\|_2$ is a normalized solution of a linear system $A\bm{x}=\bm{b}$, with $A\in \mathbb{R}^{N \times N}$ and $\bm{b}\in \mathbb{N}$. 
Since the original proposal of the QLSS~\cite{Harrow2009}, many advancements have been made to achieve better complexity\cite{Ambainis2012,Childs2017,Subasi2019,An2022,Lin2020,Costa2022}. 
Recently, a quantum algorithm has realized an optimal complexity of $\mathcal{O}(\kappa \log(1/\epsilon))$~\cite{Costa2022}, which achieves strictly linear scaling with the condition number $\kappa$ of $A$ and logarithmic scaling with the precision $\epsilon$. 
This algorithm uses the discrete adiabatic algorithm theorem, formulated based on the quantum walk operators $\{W(n/T):n\in \mathbb{N}, 0\leq n\leq T-1\}$, where $T$ is the length of walk sequence, and it is followed by improved eigenvalue filtering to provide the solution at the desired precision.

Although the optimal complexity of this algorithm has been achieved, it remains a challenging task to preserve this advantage in a full-process simulation. 
The algorithm depends on queries of the block-encoding of matrix $A$ and state preparation of vector $b$, which are resource-intensive for the current circuit model and may inhibit the quantum speed-up as claimed~\cite{Aaronson2015}. 
A potential method for verifying the efficiency of the algorithm, especially the complexity at the circuit level, is to realize the block-encoding using state-of-the-art techniques, as demonstrated in Refs.~\cite{Clader2022,Dalzell2023}.
Thus, the full-process implementation of the QLSS algorithm is necessary to assess the algorithm's capabilities. 
Furthermore, although general quantum circuit simulators such as Qiskit and Qsim can perform such simulations with well-designed operations and circuit architectures, simulating multi-qubit operations and oracles can be complex on these simulators and is not always required. 
SparQSim provides an integrated API to facilitate oracle queries during the algorithm.

\begin{figure*}[!t]
    \centering
    \begin{tikzpicture}
        \node[anchor=center] (center) at (0,0) {
            \includegraphics[width=\linewidth]{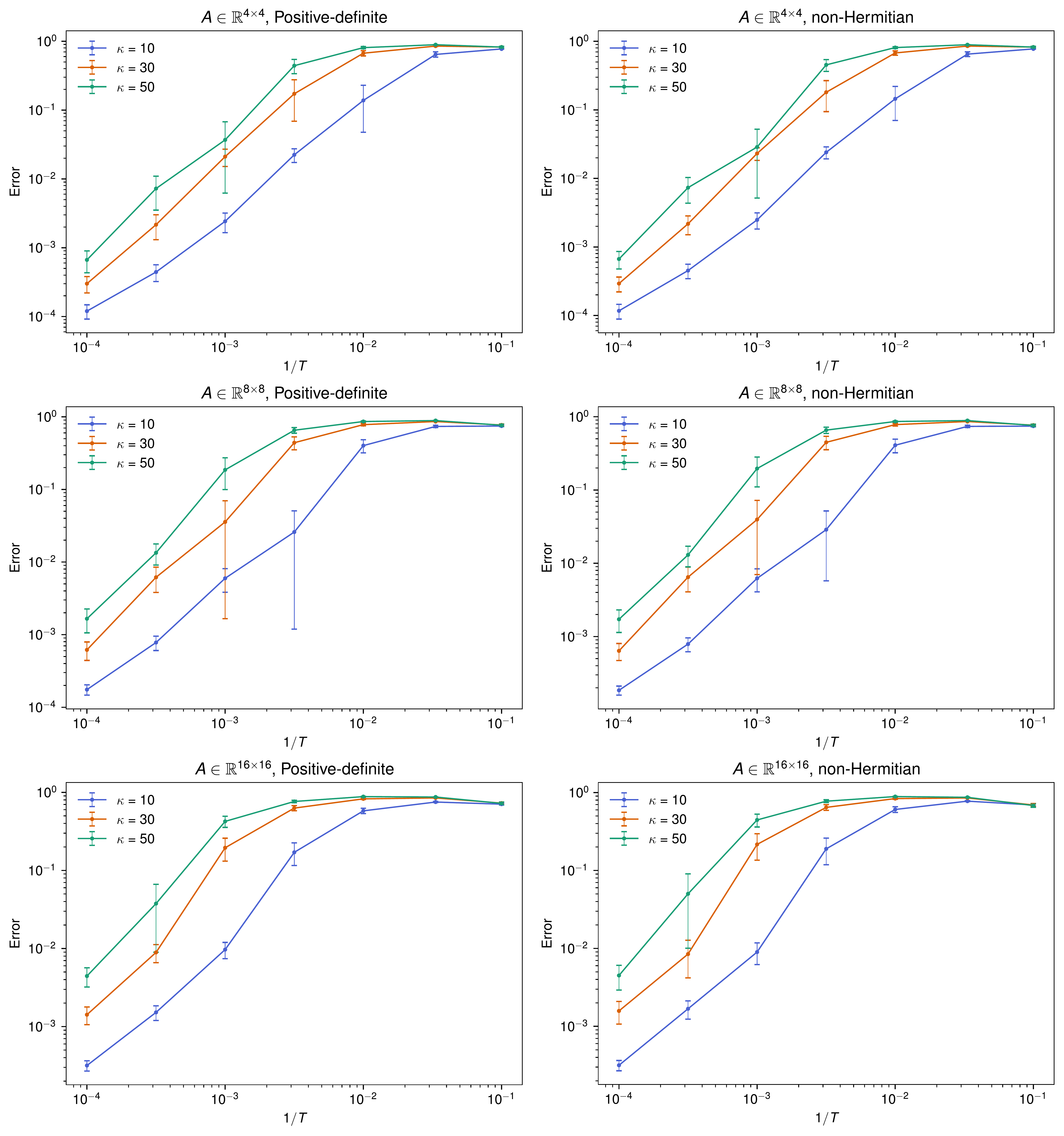}
        };
        \node[anchor=center] (a) at ($(center.north west)+(0.5cm, -0.5cm)$) {\textbf{(a)}};
        \node[anchor=center] (b) at ($(a.east)+(8.6cm, .0cm)$) {\textbf{(b)}};
        \node[anchor=center] (c) at ($(a.south)+(0.cm, -6cm)$) {\textbf{(c)}};
        \node[anchor=center] (d) at ($(b.south)+(0.cm, -6cm)$) {\textbf{(d)}};
        \node[anchor=center] (e) at ($(c.south)+(0.cm, -6cm)$) {\textbf{(e)}};
        \node[anchor=center] (f) at ($(d.south)+(0.cm, -6cm)$) {\textbf{(f)}};
    \end{tikzpicture}
    \caption{\textbf{Error Between Quantum State from Adiabatic Quantum Walks and Ideal State Versus $1/T$ for Different Linear Equations.} The error, defined as in Eq.~\ref{eq:error}, is presented for different matrix forms (positive-definite and non-Hermitian), matrix sizes ($4\times 4$, $8\times 8$, $16\times 16$), and condition numbers ($\kappa=10, 30, 50$). The error is averaged over 50 runs for each setting. Each setting is clearly indicated by the title and legends in the figure.}
    \label{fig:qlss}
\end{figure*}
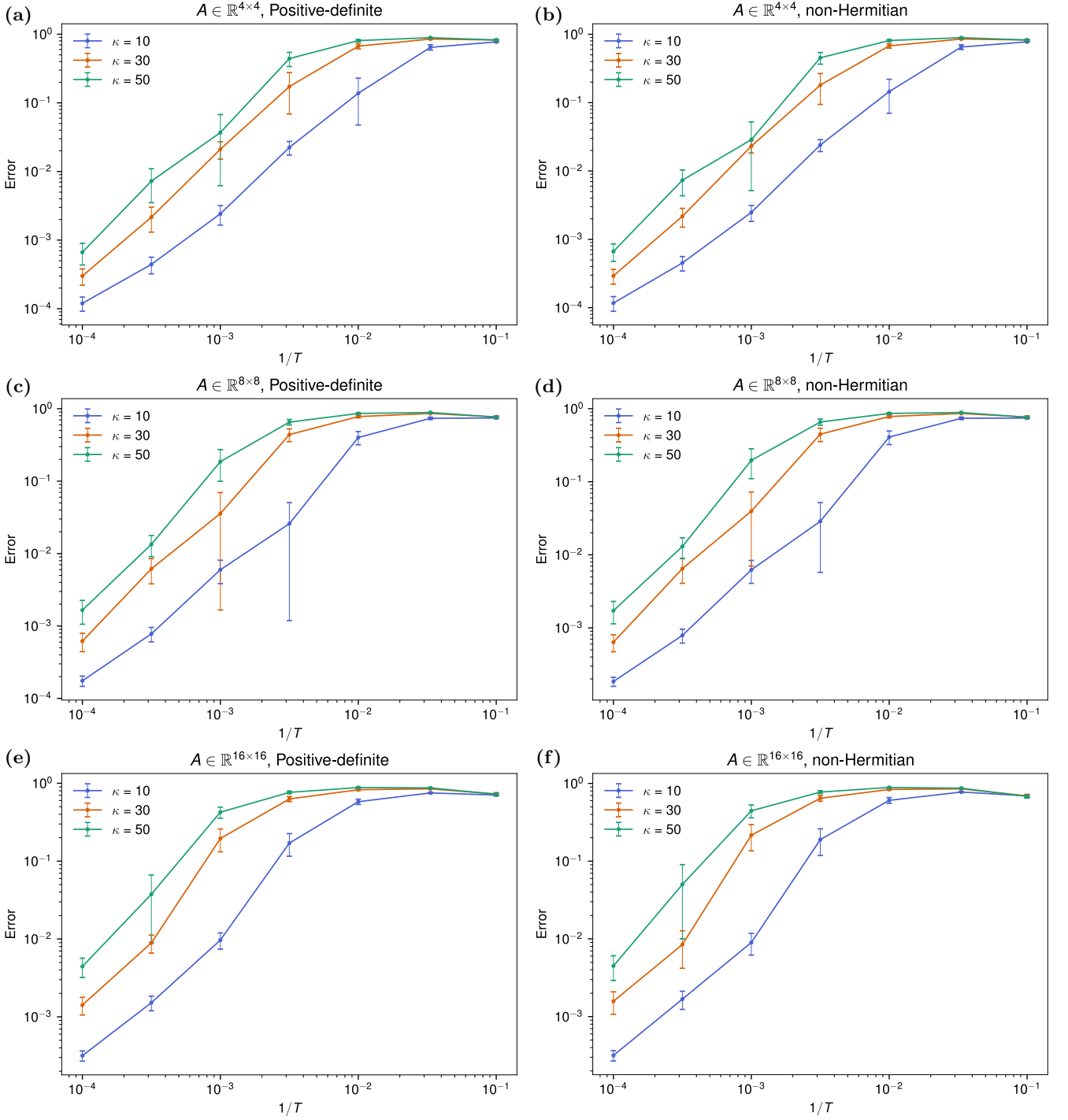

\subsubsection{The performance of the QLSS algorithm}

The discrete adiabatic theorem states that the error between the discrete evolution $U(s)$ and its ideal evolution $U_A(s)$ is bounded by the expression:
\begin{equation}
    \epsilon=\|U(s)\ket{\psi_0}-U_A(s)\ket{\psi_0}\|\leq \frac{\theta}{T},
    \label{eq:error}
\end{equation}
where $\|\cdot\|$ denotes the spectral norm and $\ket{\psi_0}$ is the initial ground state. 
The constant $\theta$ is independent of the total time $T$ but depends on the condition number $\kappa$. 
Ref.~\cite{Costa2022} shows that $\theta\approx c_1\kappa + c_2\sqrt{\kappa}$, which simplifies to $\mathcal{O}(\kappa)$.
In our work, we focus solely on the adiabatic quantum walk component of the algorithm and omit the eigenvalue filtering step. 
By varying the number of walk steps from $10^1$ to $10^4$ on a logarithmic scale, we test the algorithm on various randomly generated linear systems under different settings.
As demonstrated in Refs.~\cite{Costa2022, Costa2023}, the quantum circuits differ for positive-definite matrices and non-Hermitian matrices. 
In our simulations, we consider both cases; details of the circuits are provided in Supplementary Sec.~\ref{sec:qda}.
The matrices are generated randomly with dimensions $4\times 4$, $8\times 8$ and $16\times 16$, and condition numbers are chosen as $10$, $30$ and $50$ to introduce variability. 
Each experiment is repeated 50 times to ensure that the average error is statistically stable. 
Without loss of generality, the vector $\bm{b}$ is set to be an all-ones vector normalized to unit norm in all experiments. 
The results are shown in Fig.~\ref{fig:qlss}.

Consistent with the complexity analysis in Ref.~\cite{Costa2022}, which predicts a linear dependency on the condition number $\kappa$, our results indicate that for a fixed number of walk steps, a larger $\kappa$ leads to a larger error, as observed in Fig.~\ref{fig:qlss}. 
Moreover, for a fixed $\kappa$, the error decreases linearly as the number of walk steps increases, in agreement with the error upper bound given in Eq.~\ref{eq:error}. 
However, when the number of walk steps is reduced (i.e., when $1/T$ increases), the error becomes larger and deviates from the linear dependency, likely due to a violation of the adiabaticity condition. 
Overall, the experimental results consistently support the theoretical analysis, confirming the correctness of the full-process implementation of the QLSS algorithm.

\section{Discussion and Outlook}\label{sec:disc}

With the rapid development of quantum computing, its advantages over classical computing are becoming increasingly evident. 
Although a few quantum algorithms, such as random circuit sampling, Gaussian boson sampling and quantum simulation, demonstrate definitive complexity reductions~\cite{Larose2024}, it is essential to validate the real quantum speedup using physical quantum processors. 
In recent years, a surge of validations for these algorithms have been carried out on various platforms, including superconducting, photonic, and even annealing machines~\cite{Larose2024}.  
However, the limited size of quantum systems and the gate fidelity on NISQ devices pose significant challenges to the implementation of large-scale quantum algorithms.
Consequently, simulating quantum algorithms on classical computers is pivotal not only for validating algorithmic complexity and numerical properties but also for developing efficient and scalable simulation tools, as discussed in the introduction.

Our results indicate that SparQSim excels in scenarios with high sparsity, outperforming conventional simulators in terms of time and memory efficiency. 
However, there is substantial room for further optimization to enhance its overall performance. 
In addition, we have integrated QRAM into SparQSim to enable full-process implementations of quantum linear system solvers. 
Future work will focus on resource estimation in the fault-tolerant era and on rigorously evaluating the advantages of state-of-the-art techniques for decomposing oracles and modules into Clifford and $T$ gates.

Furthermore, realizing large-scale QRAM with fault-tolerant techniques may prove unrealistic due to the exponential overhead associated with $T$ gates. 
Thus, it is imperative to consider noise behavior in QRAM and to investigate suppression techniques to achieve better performance—an aspect that will be addressed in subsequent studies. 
We believe that SparQSim can significantly aid researchers and engineers in developing new quantum algorithms, validating quantum speedup, and gaining deeper insights into the quantum computing landscape.

\section{Acknowledgements} 
This work has been supported by the National Key Research and Development Program of China (Grant Nos. 2023YFB4502500 and 2024YFB4504100), the National Natural Science Foundation of China (Grant No. 12404564), and the Anhui Province Science and Technology Innovation (Grant No. 202423s06050001).


\bibliography{ref}

\clearpage
\setcounter{table}{0}
\renewcommand{\thetable}{S\arabic{table}}%
\setcounter{figure}{0}
\renewcommand{\thefigure}{S\arabic{figure}}%
\setcounter{section}{0}
\setcounter{equation}{0}
\renewcommand{\theequation}{S\arabic{equation}}%

\onecolumngrid

\begin{center}
{\large \bf Supplementary Information\\
SparQSim: Simulating Scalable Quantum Algorithms via Sparse Quantum State Representations}\\
\vspace{0.3cm}
\end{center}

\setcounter{page}{1}

\section{Data structure of SparQSim} \label{sec:description}

Here is the list of properties and methods of class \textbf{System} and class \textbf{StateStorage}:

\begin{table*}[h]
    \centering
    \caption{List of properties and methods built into \textbf{System} class}
    \label{tb:pm}
    \begin{tabular}{|p{3.5cm}|p{5cm}|p{8cm}|}
        \hline
        \textbf{Category} & \textbf{Property/Method} & \textbf{Description} \\ \hline
        \multirow{4}{*}{Static Properties} 
        & \verb|name_register_map| & A map that stores the name of each register and its value and data type in the register array. \\ \cline{2-3}
        & \verb|max_qubit_count| & The maximum number of system qubits during the simulation. \\ \cline{2-3}
        & \verb|max_register_count| & The maximum number of system registers during the simulation. \\ \cline{2-3}
        & \verb|max_system_size| & The maximum size of the system vector. \\ \hline
        \multirow{3}{*}{Non-Static Properties} 
        & \verb|amplitude| & The amplitude of the branch. \\ \cline{2-3}
        & \verb|registers| & The array of the registers. \\ \cline{2-3}
        & \verb|cached_hash| & The hash value for of the branch. \\ \hline
        \multirow{18}{*}{Static Methods}
        & \verb|get(id)| & Get the register with the given id. \\ \cline{2-3}
        & \verb|get(name)| & Get the register id with the given name. \\ \cline{2-3}
        & \verb|name_of(id)| & Get the register name with the given id. \\ \cline{2-3}
        & \verb|size_of(id)| & Get the register size (qubit number) with the given id. \\ \cline{2-3}
        & \verb|size_of(name)| & Get the register size (qubit number) with the given name. \\ \cline{2-3}
        & \verb|get_qubit_count()| & Get the number of qubits (all registers). \\ \cline{2-3}
        & \verb|type_of(id)| & Get the register data type with the given id. \\ \cline{2-3}
        & \verb|type_of(name)| & Get the register data type with the given name. \\ \cline{2-3}
        & \verb|status_of(id)| & Get the register status (active or inactive) with the given id. \\ \cline{2-3}
        & \verb|status_of(name)| & Get the register status (active or inactive) with the given name. \\ \cline{2-3}
        & \verb|clear()| & Clear the information of the system and registers. \\ \cline{2-3}
        & \verb|add_register(name, type, size)| & Add a new register and initialize it. \\ \cline{2-3}
        & \verb|remove_register(id)| & Remove the register with the given id. \\ \cline{2-3}
        & \verb|remove_register(name)| & Remove the register with the given name. \\ \cline{2-3}
        & \verb|get_activated_register_size()| & Get the number of activated registers. \\ \cline{2-3}
        & \verb|last_register()| & Get the last register. \\ \cline{2-3}
        & \verb|get_register_info()| & Get the information of the register with the given index. \\ \cline{2-3}
        & \verb|update_max_size()| & Update the maximum system size. \\ \hline     
    \end{tabular}
\end{table*}

\begin{table*}[h]
    \centering
    \caption{List of properties and methods built into \textbf{StateStorage} class}
    \label{tb:pm2}
    \begin{tabular}{|p{3.5cm}|p{5cm}|p{8cm}|}
        \hline
        \textbf{Category} & \textbf{Property/Method} & \textbf{Description} \\ \hline
        \multirow{1}{*}{Non-Static Properties} 
        & \verb|value| & The value of the register \\ \hline
        \multirow{4}{*}{Non-Static Methods}
        & \verb|as_bool()| & Reduce the register to a boolean value. \\ \cline{2-3}
        & \verb|val(size)| & For a safe access to the value. \\ \cline{2-3}
        & \verb|to_string()| & Convert the value to a string. \\ \cline{2-3}
        & \verb|flip(digit)| & Flip the value of specified position in the register. \\ \hline     
    \end{tabular}
\end{table*}

\section{Details of quantum operations}\label{sec:ops}
In this section, we provide detailed descriptions of the quantum operations used in the SparQSim algorithm. 
We categorize these operations into two groups: non-interference operations and interference operations.

As a representative non-interference operation, the $X$ gate is used to flip the value of a qubit on its corresponding register. 
In our implementation, we refer to this functionality as \verb|FlipBool|, which is defined as a C++ struct.

\lstset{linewidth=1.\textwidth,
    numbers=left, 
    basicstyle=\footnotesize\ttfamily,
    numberstyle=\tiny, 
    keywordstyle=\color{blue}, 
    commentstyle=\it\color[cmyk]{1,0,1,0}, 
    stringstyle=\it\color[RGB]{128,0,0},
    backgroundcolor=\color[RGB]{245,245,244},
    frame=single, 
    breaklines=true, 
    tabsize=4, 
    showstringspaces = false,
    showtabs=false,
    showspaces=false,
    language=C++,  
    morekeywords={std::string, size_t, u22_t},
    alsoletter={:},
    emph={FlipBool, operator, System, get, ConditionSatisfied,flip_digit,profiler,std::min,size},
    emphstyle=\color{magenta}
}

\begin{lstlisting}
    struct FlipBool {
		int id;
		size_t digit;
		ClassControllable
		FlipBool(std::string reg, size_t digit_): id(System::get(reg)), digit(digit_){};
		FlipBool(int id_, size_t digit_): id(id_), digit(digit_){};
		void operator()(std::vector<System>& state) const;
	};

    void FlipBool::operator()(std::vector<System>& state) const
	{
		profiler _("FlipBool");
#ifdef SINGLE_THREAD
		for (auto& s : state)
		{
			if (!ConditionSatisfied(s)) 
			continue;
			auto& reg = s.get(id);
			reg.value = flip_digit(reg.value, digit);
		}
#else
		if (state.size() < THRESHOLD)
		{
			for (auto& s : state)
			{
				if (!ConditionSatisfied(s)) 
				continue;
				auto& reg = s.get(id);
				reg.value = flip_digit(reg.value, digit);
			}
		}
		else{
			#pragma omp parallel for schedule(static)
			for (size_t chunk_start = 0; chunk_start < state.size(); chunk_start += CHUNK_SIZE)
			{
				size_t chunk_end = std::min(chunk_start + CHUNK_SIZE, state.size());
	
				for (size_t i = chunk_start; i < chunk_end; ++i)
				{
					auto& s = state[i];
	
					if (!ConditionSatisfied(s)) 
						continue;
	
					auto& reg = s.get(id);
					reg.value = flip_digit(reg.value, digit);
				}
			}		
		}	
#endif
	}
\end{lstlisting}

In our code, we use macro definitions such as \verb|SINGLE_THREAD|, \verb|THRESHOLD|, and \verb|CHUNK_SIZE| to control simulation parallelism. 
The \verb|SINGLE_THREAD| macro indicates whether to run in single-threaded or multi-threaded mode, while \verb|THRESHOLD| specifies the minimum number of systems required to activate multi-threading. 
The \verb|CHUNK_SIZE| macro defines the size of each chunk for multi-threaded processing. 
Additionally, the \verb|ClassControllable| macro manages controlled variables and their associated functions, including the \verb|ConditionSatisfied| function that checks whether a system is under control. 
The helper function \verb|flip_digit| flips the value of a specific digit in a binary number, and the \verb|profiler| function measures the execution time of a quantum operation; details of these functions are omitted here.

For interference operations, the \verb|Rot_Int| function implements a general rotation on a single qubit, following an execution strategy based on the ordered state vector: 

\begin{lstlisting}
    struct Rot_Int {
        int id;
        size_t digit;
        using angle_function_t = std::function<u22_t(size_t)>;
        size_t mask;
        u22_t mat;
        ClassControllable
        Rot_Int(std::string reg_, size_t digit_, u22_t mat): id(System::get(reg_)), digit(digit_), mat(mat), mask(1<<digit_);
        Rot_Int(int id_, size_t digit_, u22_t mat): id(id_), digit(digit_), mat(mat), mask(1<<digit_);

        void operate(size_t l, size_t r, std::vector<System>& state) const;

        static bool _is_diagonal(const u22_t& data);
        void _operate_diagonal(size_t l, size_t r,
            std::vector<System>& state, const u22_t& mat) const;

        static bool _is_off_diagonal(const u22_t& data);
        void _operate_off_diagonal(size_t l, size_t r,
            std::vector<System>& state, const u22_t& mat) const;

        void _operate_general(size_t l, size_t r,
            std::vector<System>& state, const u22_t& mat) const;

        void operate_pair(size_t zero, size_t one, std::vector<System>& state) const;
        void operate_alone_zero(size_t zero, std::vector<System>& state) const;
        void operate_alone_one(size_t one, std::vector<System>& state) const;

        void operator()(std::vector<System>& state);
    };

    void Rot_Int::operate(size_t l, size_t r, std::vector<System>& state) const
    {
        size_t n = r - l;
        constexpr size_t full_size = 2;
        size_t original_size = state.size();

        if (n == 0) return;

        if (_is_diagonal(mat))
        {
            _operate_diagonal(l, r, state, mat);
        }
        else if (_is_off_diagonal(mat))
        {
            _operate_off_diagonal(l, r, state, mat);
        }
        else
        {
            _operate_general(l, r, state, mat);
        }

    }

    void Rot_Int::operator()(std::vector<System>& state)
    {
        profiler _("Rot_Int");

        size_t current_size = state.size();
        (SortExceptBitwithOMP(id, digit))(state);
        
        auto iter_l = 0;
        auto iter_r = 1;

        while (true)
        {	
            if (iter_r == current_size)
            {
                operate(iter_l, iter_r, state);
                break;
            }
            if (!compare_equal_rot(state[iter_l], state[iter_r], id, ~mask))
            {
                operate(iter_l, iter_r, state);
                iter_l = iter_r;
                iter_r = iter_l + 1;
            }
            else
            {
                iter_r++;
            }
        }	

        ClearZero()(state);
        System::update_max_size(state.size());
    }
\end{lstlisting}
Since the sorting step, invoked as \verb|(SortExceptBitwithOMP(id, digit))(state)|, is the primary contributor to execution time, we developed an OpenMP merge sort version for cases where the state size exceeds \verb|THRESHOLD|. 
The sorted order of the state vector then determines the execution order. 
The helper function \verb|compare_equal_rot| compares two states with respect to the rotation qubit of the specified register, while \verb|ClearZero| removes zero amplitudes from the state vector. 
The \verb|System::update_max_size| function updates the maximum system size. 
Within the \verb|operate| function, we invoke \verb|_operate_diagonal|, \verb|_operate_off_diagonal|, and \verb|_operate_general| to perform the rotation operation based on the matrix \verb|mat| using different strategies.

Due to the length of the complete code, only the main components are presented here; details of the supporting functions for the rotation operation are omitted.

\section{Classical simulation of the quantum random access memory}

As an application of quantum arithmetic operations, we can implement a quantum random access memory (QRAM) using the register-level sparse state representation. QRAM is an efficient data loading module for some matrix linear algebra algorithms, among which the quantum linear system solver is one of the most important applications. As a resultl, the resource estimation of a typical quantum algorithm from end to end requires the integration of QRAM with dominant circuit part. Existing quantum simulators are welcomed to simulate QRAM explicitly with its circuit architecture. However, the simulation of QRAM is not straightforward, which requires the actual execution of quantum gates and the management of ancilla qubits. Herein, we are eager to simulate it with the register-level sparse state representation.
 
QRAM can implement a unitary $U_{\mathrm{QRAM}}$ such that
\begin{equation}
    U_{\mathrm{QRAM}}|i\rangle_A|j\rangle_D = |i\rangle_A|j\oplus d_i\rangle_D.
\end{equation}
Here, QRAM manages a vector of memory entries $\vec{d}$ and has two variables to control the input: the address length and the word length. 
The address length $n$ is the size of the address register $|\cdot\rangle_A$, and the word length is the size of data register $|\cdot\rangle_D$. 
The word length is also the effective size of a memory entry $d_i$. 
When the noise is not considered, the simulation of QRAM is to access the classical memory with the content of the address register.

The simulation of QRAM consists of several ingredients: the initialization of the registers inlcuding address register, data register and ancilla register; the QRAMLoad operation and its inverse. The relationship between the QRAM and block-encoding of a matrix will be discussed in the next section.

\section{Implementation of Quantum Linear-System Solver with QRAM}\label{sec:qda}

\begin{figure*}[!t]
    \centering
    \begin{tikzpicture}
        \node[anchor=center] (a) at (0,0) {
            \includegraphics[width=\linewidth]{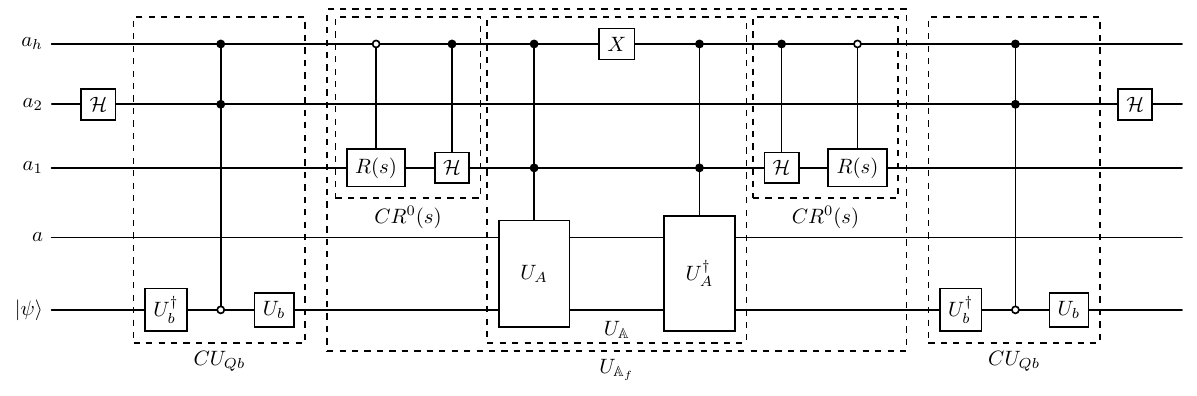}
        };
        \node[anchor=center] (b) at ($(a.south)-(0, 3cm)$) {
            \includegraphics[width=\linewidth]{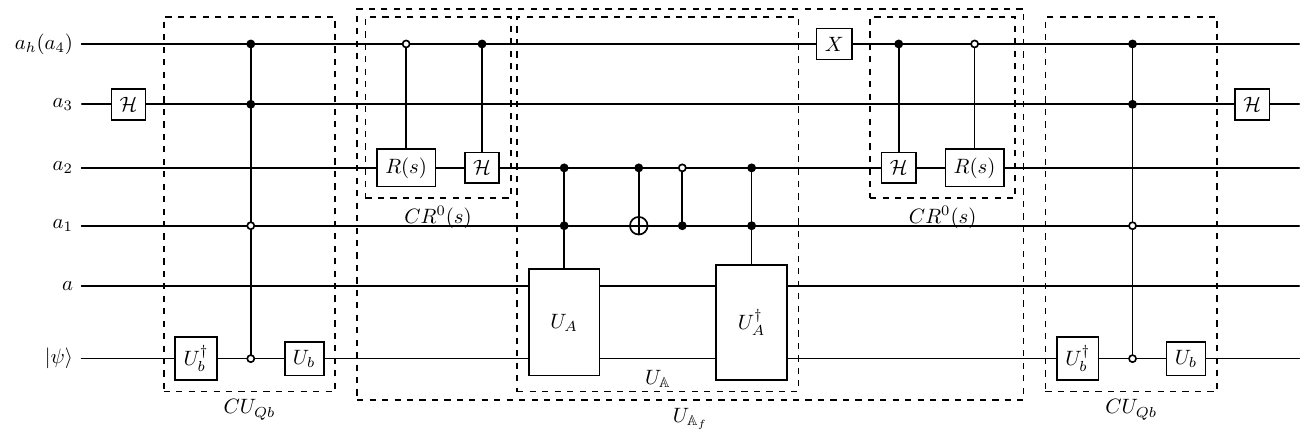}
        };
        \node[anchor=center] (c) at ($(b.south)-(0, 1cm)$) {
            \includegraphics[width=\linewidth]{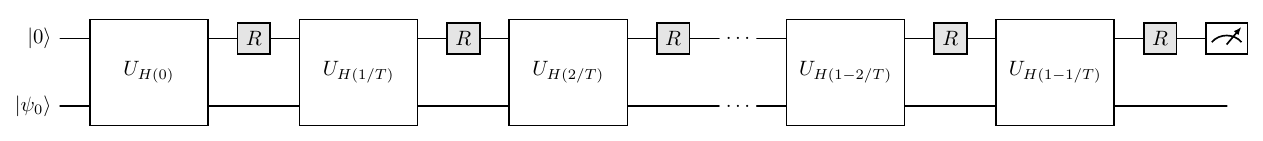}
        };
        \node[anchor=north west] at ($(a.north west)-(0.2cm,0.2cm)$) {\textbf{(a)}};
        \node[anchor=north west] at ($(b.north west)-(0.2cm,0.2cm)$) {\textbf{(b)}};
        \node[anchor=north west] at ($(c.north west)-(0.2cm,0.2cm)$) {\textbf{(c)}};
    \end{tikzpicture}
    \caption{(a) Block-encoding of $H(s)$ for a positive-definite matrix. (b) Block-encoding of $H(s)$ for a non-Hermitian matrix. (c) Quantum adiabatic evolution with qubitized walks.}
    \label{fig:qc1}
\end{figure*}

Adiabatic quantum computing requires two different Hamiltonians, $H_0$ and $H_1$, which have ground states $\ket{\psi_0}$ and $\ket{\psi_1}$, respectively. 
The goal is to prepare $\ket{\psi_1}$ from the easily prepared initial state $\ket{\psi_0}$ by following an adiabatic path. 
This path is defined by a family of Hamiltonians $H(s)=[1-f(s)]H_0+f(s)H_1$, with a schedule function $f(s): [0,1] \to [0,1]$. 
In the context of the quantum linear-system problem, the vector $\bm{b}$ is encoded as the initial state and the solution $\bm{x}$ as the target state. 
Accordingly, the Hamiltonian $H(s)$ can be explicitly implemented by combining $H_0$ and $H_1$. 
The corresponding Hamiltonians and quantum states are listed in the following table:

{
    \renewcommand{\arraystretch}{1.2} 
    \setlength{\extrarowheight}{2pt}   
    \begin{longtable}{c|c|c}
        \hline
         & Positive-definite & non-Hermitian \\ 
         \hline
         $H_0$
         &  $H_0=\begin{pmatrix}
            0& Q_b\\ Q_b & 0 \end{pmatrix}$            
         &  $H_0=\begin{pmatrix}
            0& (\sigma_z\otimes I)Q_{\bm{\mathrm{b}}}\\ Q_{\bm{\mathrm{b}}}(\sigma_z\otimes I) & 0 \end{pmatrix}$\\ \hline
         $H_1$
         &  $H_1=\begin{pmatrix}
            0& AQ_b\\ Q_bA & 0 \end{pmatrix}$ 
         &  $H_1=\begin{pmatrix}
            0& (\sigma_z\otimes A)Q_{\bm{\mathrm{b}}}\\ Q_{\bm{\mathrm{b}}}(\sigma_z\otimes A) & 0 \end{pmatrix}$ \\ \hline
         $H(s)$
         &  $H(s)=\begin{pmatrix} 
            0 & A(f)Q_b \\ Q_b A(f) & 0
            \end{pmatrix}$, 
            $A(f)=(1-f)I + fA$
         &  $H(s)=\begin{pmatrix} 
            0 & A(f)Q_{\bm{\mathrm{b}}} \\ Q_{\bm{\mathrm{b}}} A(f) & 0
            \end{pmatrix}$, 
            $A(f)=\begin{pmatrix}
            (1-f)I & fA \\ fA^\dagger & (1-f)I
            \end{pmatrix}$ \\ 
            \hline
         $\ket{\psi_0}$
         &  $\begin{pmatrix}b \\ 0\end{pmatrix}$, or $\ket{0}_{a_h}\ket{b}$
         &  $\begin{pmatrix}\bm{\mathrm{b}} \\ 0\end{pmatrix}$, or $\ket{0}_{a_h}\ket{0}_{a_1}\ket{b}$ \\ 
         \hline
         $\ket{\psi_1}$ 
         &  $\begin{pmatrix}A^{-1}b \\ 0\end{pmatrix}$, or $\ket{0}_{a_h}\ket{A^{-1}b}$
         &  $\begin{pmatrix}\bm{\mathrm{A}}^{-1} \bm{\mathrm{b}} \\ 0\end{pmatrix}$, or $\ket{0}_{a_h}\ket{0}_{a_1}\ket{A^{-1}b}$\\ 
         \hline
         Remarks
         &$Q_b=I-\ket{b}\bra{b}$  & $Q_{\bm{\mathrm{b}}}=I-\ket{\bm{\mathrm{b}}}\bra{\bm{\mathrm{b}}}$, $\ket{\bm{\mathrm{b}}}=\ket{0}_{a_1}\ket{b}$, $\bm{\mathrm{A}}=\begin{pmatrix}
            0 & A \\ A^\dagger & 0
         \end{pmatrix}$   \\  \hline
    \end{longtable}

}

Since the circuit model requires unitary operations, the operations such as $Q_b$, $Q_{\bm{b}}$, and $A(f)$ must be implemented using quantum gates or oracles. Block-encoding techniques are used to embed $A$ into a block matrix $U_A$ such that the top-left block of $U_A$ equals $A/\alpha$ for some normalizing constant $\alpha \geq \|A\|$; that is,  $\alpha (\bra{0}^{\otimes a} \otimes I) U_A (\ket{0}^{\otimes a} \otimes I)=A$. 
In other words, with a query to the oracle $U_A$, one can readily construct the Hamiltonian term corresponding to $A$. The operations $Q_b$ and $Q_{\bm{b}}$ are implemented via a state preparation oracle $U_b: \ket{0} \mapsto \ket{b}$. With these oracles, the full block-encoding of $H(s)$, denoted as $U_{H(s)}$, is obtained using the quantum circuits shown in Fig.~\ref{fig:qc1}(a) and (b), thereby transforming the problem into one of discrete quantum walks, as depicted in Fig.~\ref{fig:qc1}(c). In Fig.~\ref{fig:qc1}(c), a single operation for block-encoding $U_{H(s)}$ is represented by a large block, and $R$ denotes the reflection on the ancilla qubits, defined as 
\begin{equation}
    R=i(2\ket{0}\bra{0}_{\text{anc}}-I_{\text{anc}})\otimes I.
\end{equation}

\begin{figure}[h]
    \includegraphics[width=\linewidth]{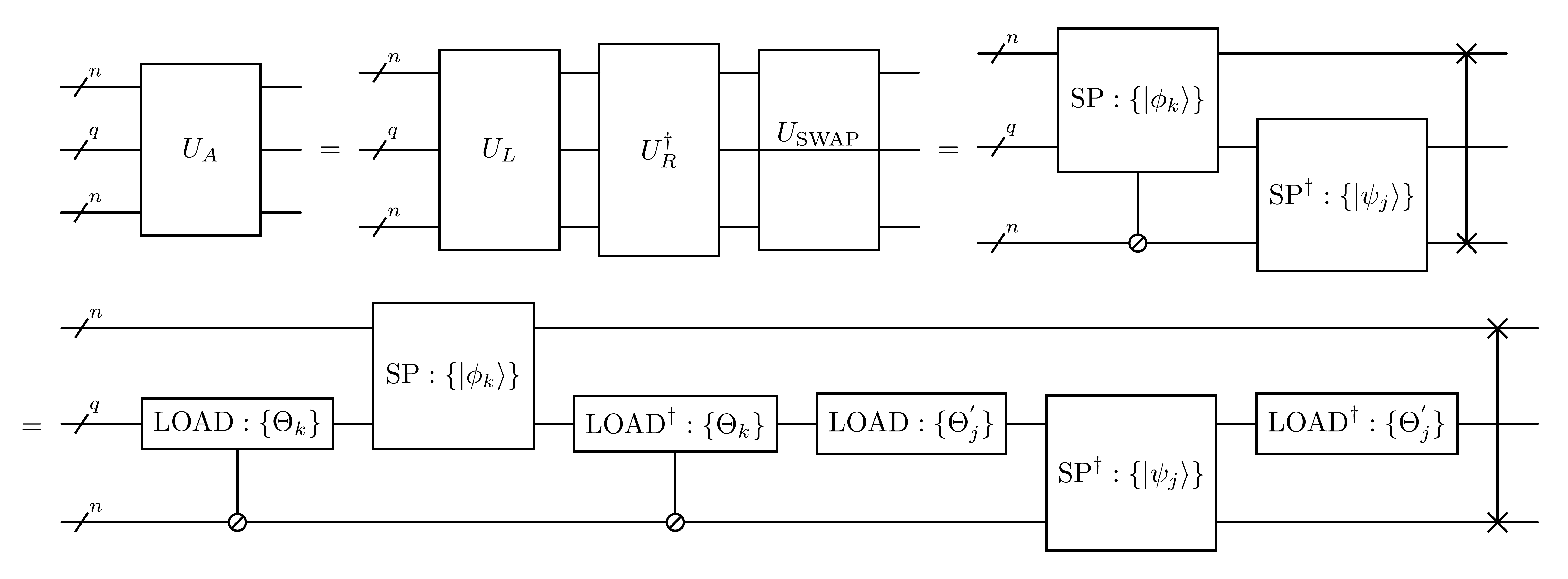}
    \caption{\textbf{Reduction from block-encoding to controlled state preparation.} }
    \label{fig:qc4}
\end{figure}

The implementation of the block-encoding of $H(s)$ is based on queries to the operations $U_A$ and $U_b$, which are, respectively, the block-encoding of the matrix $A$ and the state preparation oracle for $\ket{b}$. Since block-encoding can be further reduced to controlled state preparation as shown in \cite{Clader2022}, we describe here only the implementation details of $U_A$; the implementation of $U_b$ is a straightforward special case. To begin with, $U_A$ can be decomposed into three operations:

\begin{equation}
    U_A = U_{\text{SWAP}}U_R^\dagger U_L,
\end{equation}
where 
\begin{equation}
    \begin{aligned}
        U_L\ket{k}_n\ket{0}_q\ket{0}_n =& \ket{k}_n \ket{0}_q \ket{\phi_k}_n, \\
        U_R\ket{0}_n\ket{0}_q\ket{j}_n =& \ket{\psi}_n \ket{0}_q \ket{j}_n, \\
        U_{\text{SWAP}}\ket{i}_n\ket{0}_q\ket{j}_n =& \ket{j}_q \ket{0}_n \ket{i}_q.
    \end{aligned}
\end{equation}

Here, the states $\ket{\phi_k}_n$ and $\ket{\psi}_n$ are defined as:
\begin{equation}
    \begin{aligned}
        \ket{\phi_k}_n = & \sum_j\frac{A_{jk}}{\|A_{.,k}\|}\ket{j}_n \\
        \ket{\psi}_n = & \sum_k\frac{\|A_{.,k}\|}{\|A\|}\ket{k}_n.
    \end{aligned}
\end{equation}
where $\|A_{.,k}\|$ represents the column norm of $A$ for the $k$-th column, and $\|A\|$ is the Frobenius norm of $A$.
As shown in Fig.~\ref{fig:qc4}, the middle register with $q$ qubits is dedicated to QRAM, while the top and bottom registers with $n$ qubits correspond to the target register and the ancilla register (denoted as $a$ in Fig.~\ref{fig:qc1}(a) and (b)), respectively. Pre-computed rotation angles, organized in a binary tree structure as $\Theta_k$ and $\Theta_j$ for $\ket{\phi_k}_n$ and $\ket{\psi}_n$, are stored in classical memory and queried by QRAM, as illustrated in Fig.~\ref{fig:qc4}. With this setup, one can verify that 
\begin{equation}
    \bra{i}_n\bra{0}_q\bra{0}_n U_A \ket{j}_n\ket{0}_q\ket{0}_n=A_{ij}/\|A\|.
\end{equation}

\end{document}